\documentclass{elsart}

\usepackage{amssymb,amsmath}

\usepackage[dvips]{graphicx}

\newcommand{\comment}[1]{}
\begin{document}

\begin{frontmatter}
\title{The spin-double refraction in two-dimensional
electron gas}

\author{D. Bercioux\thanksref{now}\corauthref{cor}}
\corauth[cor]{Corresponding author.}
\ead{bercioux@na.infn.it}
\thanks[now]{Present address: Institut f{\"u}r Theoretische Physik,
Universit{\"a}t Regensburg, D-93040, Germany}
and
\author{V. Marigliano Ramaglia}

\address{{\it Coherentia}-INFM and Dipartimento di Scienze
Fisiche Universit\`a degli Studi {\it Federico II}, Napoli,  I-80126,
Italy} 

\begin{abstract}
We briefly review the phenomenon of the spin-double refraction that
originates at an interface separating a two-dimensional electron gas
with Rashba spin-orbit coupling from a one without.
We demonstrate how this phenomenon in semiconductor
heterostructures can produce and control a spin-polarized current
without ferromagnetic leads. 
\end{abstract}

\end{frontmatter}

Spintronics is a multidisciplinary field whose central subject is the
active control and manipulation of spin degrees of freedom in
solid-state systems~\cite{awschalom,zutic}. The attempts to realize
the spin Field-Effect Transistor (spin-FET) proposed by Datta and
Das~\cite{Datta}, based on Rashba Spin-Orbit (SO)
coupling~\cite{Rashba}, have been unsuccessful because of the
difficulty of inject spin-polarized carriers from a ferromagnetic
metal into a semiconductor~\cite{Schmidt}.  One of the way to solve
the problem of the conductivity mismatch at the
ferromagnet-semiconductor interface is to use dilute magnetic
semiconductor as input source~\cite{pala}.

We have proposed~\cite{Marigliano} an alternative way to overcome the
problem of the conductivity mismatch. In this work we consider the
phenomenon of the spin-double refraction. A similar effect has been
successively presented by Khodas \emph{et al.}~\cite{khodas}.
This is observed in two-Dimensional Electron Gas (2DEG) when electrons
are injected with an angle out of the normal on an interface
separating a region without Rashba SO coupling (NR zone) from a region
with it (R zone). The behavior of electron spin in such scattering is
analogous to the polarization of the light in a biaxial crystal: the
incident ray splits, within the crystal, in two rays (ordinary and
extraordinary) whose polarizations are orthogonal.

%
%
\begin{figure}
	\centering \includegraphics[width=2.2in]{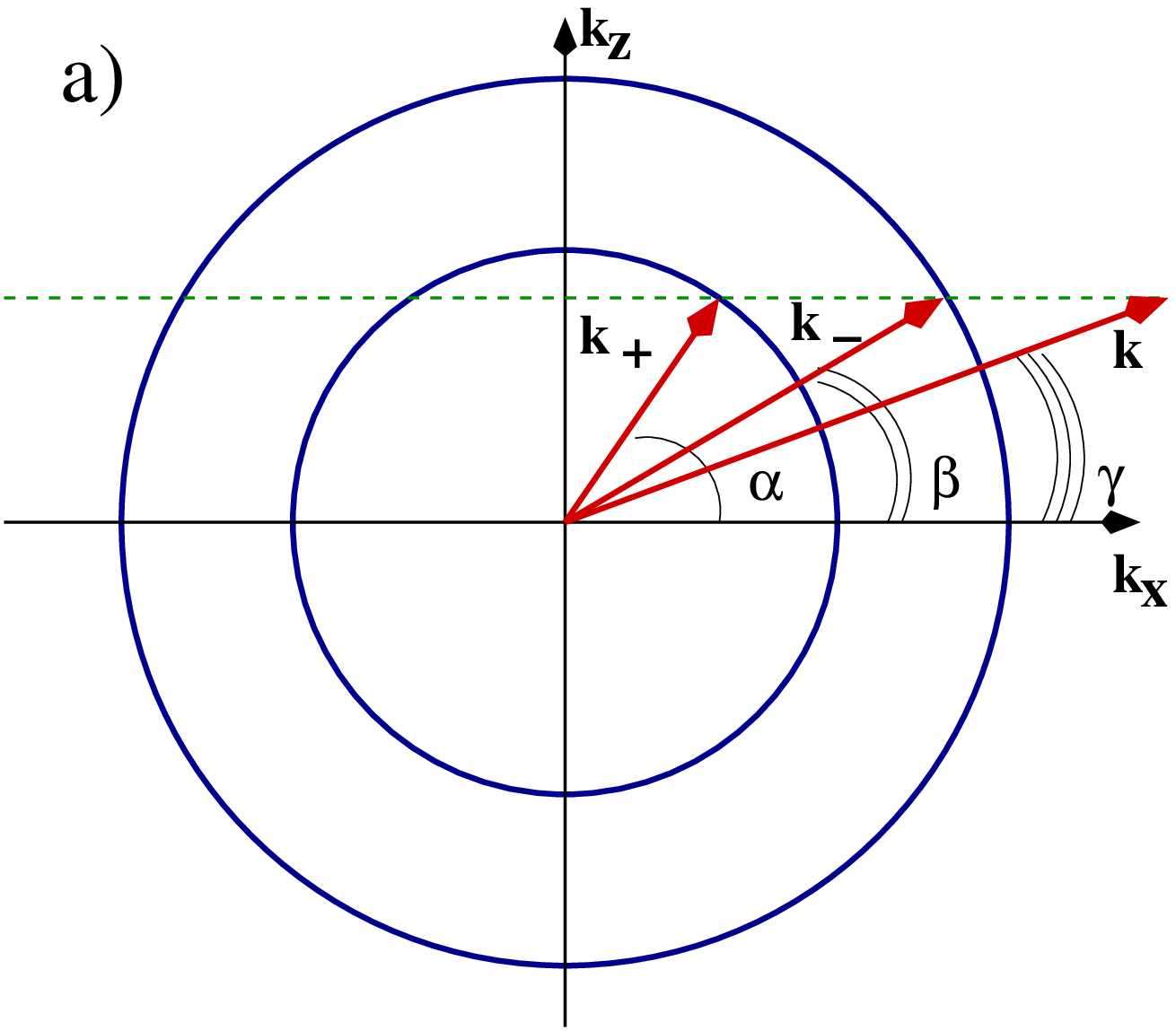}
	\includegraphics[width=3in]{figure2.eps} \caption{Panel a):
	the vectors $\vec{k}_{+}$, $\vec{k}_{-}$ and $\vec{k}$ in
	$k$-space and the angles $\alpha$, $\beta $ and $\gamma$ that
	they form with $x$ direction normal to the interface. The two
	circles are the Fermi contour at a fixed energy. Panel b):
	the conductance for polarized electrons $\mathcal{G}$ as a
	function of the spin polarization $\delta $ of the incoming
	electrons. \comment{The parameters are $k=10k_\text{SO}$,
	$\mu=0.1$ and $u=1$.} The value $\delta =0$ corresponds to
	spin up and $\delta =\pi /2$ to spin down. \label{fig:one}}
\end{figure}
%
%

The Hamiltonian describing the hybrid system is
%
%
\begin{equation}\label{eq:one}
\mathcal{H}_{\text{NR-R}} = \vec{p}\frac{1}{2 m(x)}
\vec{p}+\frac{k_\text{SO}(x) m(x)}{\hbar^2} 
\left( \sigma_z p_x -\sigma_x p_z\right) -i \sigma_z \frac{1}{2}
\frac{\partial 
k_\text{SO}(x)}{\partial x}+ U \delta(x).
\end{equation}
%
%
The second term is the Rashba SO coupling. The presence of the
interface is introduced by the mass and the strength of SO coupling
that are piecewise constant
%
%
\begin{equation}\label{eq:two}
\frac{1}{m}(x) = \frac{\vartheta(-x)}{m_\text{NR}} +
\frac{\vartheta(x)}{m_\text{R}} ~~\text{and}~~
k_\text{SO}(x) = k_\text{SO}~ \vartheta(x) ,
\end{equation}
%
%
where $\vartheta(x)$ is the step function. The third term in
(\ref{eq:one}) is needed to get an hermitian operator
$\mathcal{H}_{\text{NR-R}}$ and the fourth term regulates the
transparency of the interface. Let us consider an electron injected
from the NR zone on the interface with momentum $k$ and angle
$\gamma$.  The electron wave function in the R zone is given by the
superposition of the split eigenstates of the R zone, whose directions
are fixed by the conservation of the momentum parallel to the
interface
%
%
\begin{equation}\label{eq:three}
k_+ \sin \alpha = k_- \sin \beta = k \sin k
\end{equation}
%
%
where $k_\pm = \sqrt{\mu k^2 +k_\text{SO}^2} \mp k_\text{SO}$ is the
momentum relative to the two eigenstates (see Fig.~\ref{fig:one} Panel
a) and $\mu=m_\text{N}/ m_\text{NR}$. The two eigenstates of the R
zone are characterized by two limit angles
%
%
\begin{equation}\label{eq:four}
\gamma_0 = \arcsin \frac{k_+}{k} ~~~\text{and}~~~ \gamma_1 = \arcsin
\frac{k_-}{k}>\gamma_0
\end{equation}
%
%
over which they become non propagative waves.  The analysis of the
transmission probability of the hybrid system shows that in the case
of normal incidence the interface is not able to distinguish spin up
and spin down electrons. Instead, the oblique scattering, due to the
spin-double refraction, gives rise to an output spin up probability
different from the spin down probability~\cite{Marigliano}. This
effect of polarization survives also when all the injection angles are
taken into account ($-\gamma_1 < \gamma < \gamma_1$). This is
equivalent to consider the transmission through a wide point contact
devices. In the Fig.~\ref{fig:one} Panel b) is shown the behavior
of the conductance $\mathcal{G}$ of this system as function of the spin
state $\delta$. It is evident that the conductance of the single
interface when the incoming spin is up ($\delta=0,\pi$) is different
from the conductance when the incoming spin state is down
($\delta=\pi/2$). This effect is a direct consequence of the
spin-double refraction at the interface that change the spin state
when the electron comes into the R zone.

\begin{figure}
	\centering
	\includegraphics[width=2.5in]{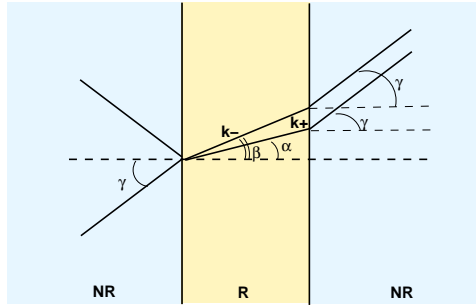}
	\caption{Schematic illustration of the proposed devices. The
	two-dimensional electron gas is divided in three
	region.  In central region a Rashba spin-orbit coupling is
	present. \label{fig:two}}
\end{figure}

\begin{figure}
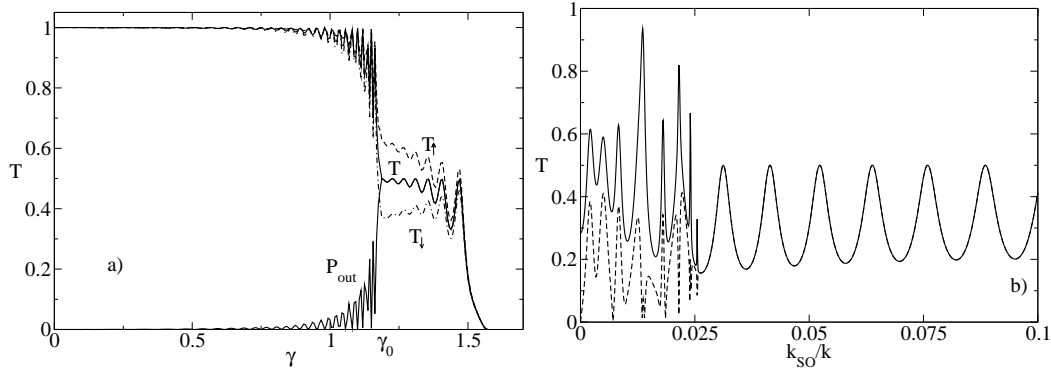

	\centering \includegraphics[width=2.7in]{figure4.eps}
	\includegraphics[width=2.7in]{figure5.eps} \caption{Panel a):
	The transmission coefficients $T_{\downarrow }\left(
	\gamma\right)$, $T\left( \gamma \right)$, $T_{\uparrow }\left(
	\gamma \right) $ as functions of the incidence angle $\gamma $
	(in radians). Panel b): The transmission coefficient for
	unpolarized electrons when $\gamma =1.25$. The threshold
	$\overline{k}_{SO}$ is $0.025k$. The dotted curve gives the
	output spin polarization, the length of the Rashba zone is
	$L=283$~nm. \label{fig:three}}
\end{figure}

This phenomenon can be used to realize an spin-FET without
ferromagnetic contacts~\cite{marigliano:2004}. The system is realized
in a 2DEG where a region in which the Rashba SO coupling is present
(see Fig.\ref{fig:two}).
%
%
%
%
The source and the drain could be realized using n$^+$-semiconductors.
The novel feature of our setup is the transmission double step shown
in Fig.~\ref{fig:three} Panel a), that is accompanied by the
appearance of a spin polarization. This is due to the spin-double
refraction, when $\gamma$ goes over $\gamma_0$ only one mode of the R
zone can reach the second interface and the transmission coefficients
tend to halve themself.
%
%
%
%
The dependence of the limit angle $\gamma_0$ on the SO
strength $k_\text{SO}$ suggests the possibility to build up a spin-FET
operating on spin unpolarized electrons injected in the R region. The
electrons emerge in the NR drain region partially polarized with a
polarization controlled by a gate electrode via the SO
interaction. There is a SO strength $\overline{k}_\text{SO}$ at which
$\gamma =\gamma_0$. The Fig.~\ref{fig:three} Panel b) shows that the
transmission coefficient exhibits irregular Fabry-Perot oscillations
below $\overline{k}_\text{SO}$, whereas for $k_\text{SO}>
\overline{k}_\text{SO}$ the oscillations become regular 
with maxima equal to $1/2$. The threshold $\overline{k}_\text{SO}$ is
determined by the parameters of the Hamiltonian of the hybrid
system~\cite{marigliano:2004}. When $k_\text{SO}>
\overline{k}_\text{SO}$ the polarization of electrons in the NR drain
$P_{\text{out}}$ is equal to $T$. We get a source of a spin polarized
current controlling $k_\text{SO}$ with a gate.  We stress that a
modulation of the output current can be obtained with a
spin-unpolarized input current, whereas in the original Datta and
Das~\cite{Datta} proposal the current oscillation stems out from the
difference of phase accumulated along a path in the Rashba region by
the two spin propagating modes. In our system the current modulation
and the spin polarization appear when {\it only one mode} propagates
through the Rashba barrier.

\end{document}